\documentclass[aip,rsi,preprint,graphicx]{revtex4-1}

\usepackage{colortbl}
\usepackage[table]{xcolor}
\usepackage{tabularx}
\usepackage{amsmath}
\usepackage[latin1]{inputenc} \usepackage{graphicx}
\usepackage{color} 

\definecolor{darkblue}{rgb}{0,0,0.5}
\definecolor{lila}{rgb}{0.3,0,0.3}
\definecolor{turq}{rgb}{0,0.1,0.4}
\usepackage[colorlinks=true, linkcolor=darkblue, 
  filecolor=red, citecolor=turq, 
]{hyperref}

\begin{document}

\title{Microscopic diamond Solid-Immersion-Lenses fabricated around single
defect ceneters by focussed ion beam milling}

\author{Mohammad Jamali}
\affiliation{3. Physikalisches Institut and Stuttgart Research Center of Photonic Engineering (SCoPE), Universit\"at Stuttgart, Pfaffenwaldring 57, Stuttgart D-70569, Germany}
\author{Ilja Gerhardt}
\affiliation{3. Physikalisches Institut and Stuttgart Research Center of Photonic Engineering (SCoPE), Universit\"at Stuttgart, Pfaffenwaldring 57, Stuttgart D-70569, Germany}
\affiliation{Max Planck Institute for Solid State Research, Heisenbergstra\ss e 1, D-70569 Stuttgart, Germany}
\author{Mohammad Rezai}
\affiliation{3. Physikalisches Institut and Stuttgart Research Center of Photonic Engineering (SCoPE), Universit\"at Stuttgart, Pfaffenwaldring 57, Stuttgart D-70569, Germany}
\author{Karsten Frenner}
\affiliation{Institute for technical optics (ITO), Universit\"at Stuttgart, Pfaffenwaldring 7, Stuttgart D-70569, Germany}
\author{Helmut Fedder}
\affiliation{3. Physikalisches Institut and Stuttgart Research Center of Photonic Engineering (SCoPE), Universit\"at Stuttgart, Pfaffenwaldring 57, Stuttgart D-70569, Germany}
\email{helmut.fedder@gmail.com}
\author{J\"org Wrachtrup}
\affiliation{3. Physikalisches Institut and Stuttgart Research Center of Photonic Engineering (SCoPE), Universit\"at Stuttgart, Pfaffenwaldring 57, Stuttgart D-70569, Germany}
\affiliation{Max Planck Institute for Solid State Research, Heisenbergstra\ss e 1, D-70569 Stuttgart, Germany}

\date{\today}

\begin{abstract}
Recent efforts to define microscopic solid-immersion-lenses (SIL) by focused
ion beam milling into diamond substrates that are registered to a preselected
single photon emitter are summarized.
We show how we determine the position of a single emitter with at least 100 nm
lateral and 500 nm axial accuracy, and how the milling procedure is optimized.
The characteristics of a single emitter, a Nitrogen Vacancy (NV)
center in diamond, are measured before and after producing the SIL and
compared with each other. A count rate of 1.0 million counts per second is
achieved with a $[111]$ oriented NV center.
\end{abstract}

\maketitle

\section{Introduction}
Single optical emitters embedded in solid state materials are on the
research horizon for more than twenty years~\cite{moerner_prl_1989}. Not only their single photon
emission, but also their nanoscopic size, and their properties as single optical
and magnetic qubits allow for various quantum optics, quantum information
and sensing experiments. Among them, color centers in diamond especially
the negatively charged nitrogen vacancy (NV) center emerges as one of the promising candidates.
It has a strong optical transition at 637 nm and local electron and nuclear
spins with long spin coherence times even at room temperature that are suitable
for quantum memories.
Therefore, it has been used in different applications, such as quantum registers, magnetic field
sensors and diamond-based single photon sources
~\cite{jelezko_pssa_2006,dutt_science_2007,Brouri_opt_2000,kurtsiefer_rpl_2000,
balasubramanian_nature_2008}.

Generally, a solid state environment allows for high collection
efficiencies~\cite{koyama_apl_1999,lee_np_2011}. The acceptance angle into the
collection optics can be very high, and high refractive index allows for a small
focus size. Various methods to increase collection efficiency have been proposed
and experimentally studied~\cite{barnes_epjd_2002}. Among them, thin
layers~\cite{lee_np_2011}, pillar structures~\cite{babinec_nn_2010}, and solid
immersion lenses (SILs) have been explored~\cite{mansfield_apl_1990}. A solid
immersion lens increases the collection efficiency from a single emitter, by
circumventing refraction from interfaces and thereby increasing the numerical
aperture. Different geometries, namely the hemispherical SIL and the Weierstrass
SIL have been researched~\cite{barnes_epjd_2002}. The later is not optimal for spectrally
broad emitters, since the shape results in strong chromatic abberations.
Hemispherical solid immersion lenses have been successfully used in single
emitter studies, e.g. with
single molecules~\cite{wrigge_np_2008}, quantum dots~\cite{vamivakas_nl_2007}
and single defects in solids~\cite{robledo_n_2011,Waldherr2014,Kolesov2013}.

Using a SIL for luminescent defects in diamond is especially interesting,
since diamond has one of the highest refractive indices in the visible range
($n_d=2.42$).
The refractive index difference at the diamond-air interface causes strong
refraction and total internal reflection (critical angle $24^\circ$) for the
emitted light. Therefore, light emitted by a defect cannot be efficiently
collected. Also for ideal spin properties, the NV defects under study should be embedded
deeply in substrate~\cite{Naydenov2011}. Hence, it is important to fabricate special
optical structures to enhance collection efficiency.

The approach of producing SILs directly on diamond started
recently~\cite{siyushev_apl_2010}. Two approaches have been established: To
produce macroscopic half-spheres, with length scales of
milimeters~\cite{siyushev_apl_2010}, and to produce microscopic SILs in the
order of serval micrometers~\cite{marseglia_apl_2011,robledo_n_2011}.
Macroscopic SILs were produced by laser and mechanical processing from small
single crystalline CVD diamonds, which are overgrown on high quality high
temperature high pressure (HPHT) grown diamond substrate. Microscopic SILs can
be produced by focused ion beam (FIB) milling.

This paper outlines the microscopic manufacturing process for SILs. We describe
the pathway to manufacture a solid immersion lens: first, a single emitter is
optically located and characterized. Afterwards a SIL is manufactured around it.
This is achieved by FIB milling. We compare different milling
strategies and present the one that is optimal to produce SILs closest to the
desired hemispherical shape and with least milling residuals.

\section{Locating a single emitter}
Before producing a SIL around a single emitter, we have to locate the
emitter in all three spatial directions.
Since these steps, the characterization
under confocal microscopy and milling in the FIB machine, are performed in two
different setups, it is required to introduce suitable marker structures
that are visible in both microscopes and to
which the position of the emitter and the SIL are referenced.
This is required for lateral localization. For the
localization in depth, we do not require a retrievable structure, since the
surface serves as a reference.

We first discuss the lateral localization of the emitter: The localization
accuracy of a single NV should be in the range of the field of view of the SIL.
The field of view diameter, $d_{FOV}$, follows from a quater-wave
criterion \cite{baba_jap_1999} and is proportional to the square root of the SIL
radius, $r$.
\begin{equation}
    d_{FOV}<\sqrt{\frac{2r\lambda}{n(n-1)}},
    \label{eq:field}
\end{equation}
where $\lambda$ is the optical wavelength and $n$ is the refractive index of
the SIL.
For a $4~\mu$m SIL in diamond and for $\lambda=532$~nm this is about 1~$\mu$m.
Consequently, the ability to locate an emitter, has to be significantly better
then 1~$\mu$m, both in the FIB machine and the optical microscope.

\begin{figure}
  \includegraphics[width=70mm]{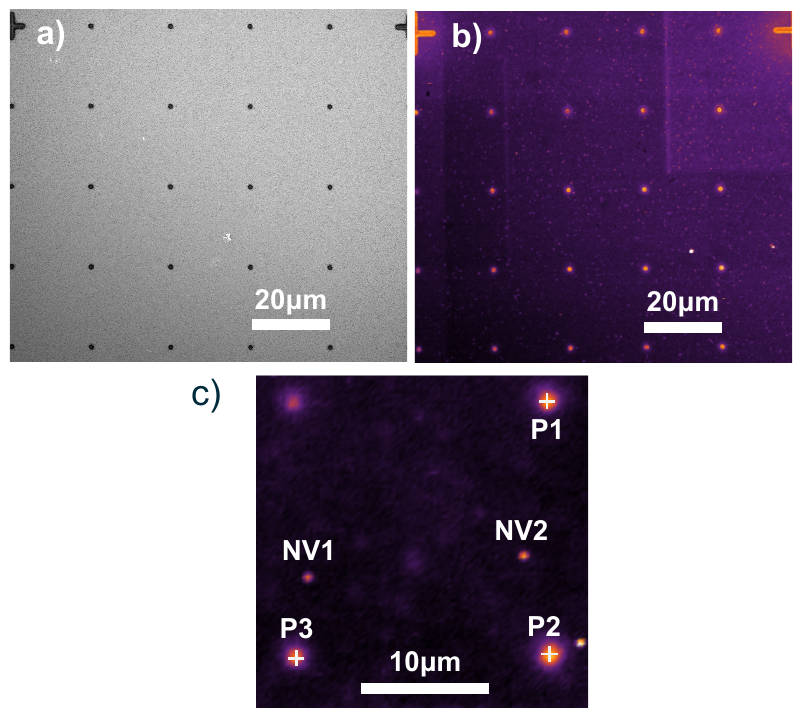}
  \caption{a) SEM image of focused ion beam generated markers in a rectangular
  pattern of 20$\times$20$~\mu$m. Pixel size: 120~nm. b) Optical image of the
  same markers, acquired in a confocal microscope. Pixel size: 200~nm. c)
  Confocal image of Sample showing two NVs and the markers around them. Note
  that the markers appear larger since the focal plane coincides with the NVs
  such that the markers are out of focus.}
  \label{fig:marker}
\end{figure}

We locate the emitter by
measuring its relative position against three marker points in the confocal
microscope, further we locate these points under the FIB and calculate the
actual position of the NV. For this purpose we mill a rectangular pattern of
cylindrical holes (diameter $~260\pm20$~nm, depth $~$500~nm, pitch $20~\mu$m,
current 0.92 nA) into the sample with the FIB. These markers are well visible in
the confocal fluorescence microscope (Fig.\ref{fig:marker}a,b). It is
currently not clear what is at the origin of their fluorescence.
The fluorescence could either stem from
graphitized material, implanted gallium or dirt contained in the immersion oil
that is trapped inside the holes.
Single emitters are laterally located by imaging the sample at the target
emitter depth below the surface with a
home-made confocal microscope~(Fig.\ref{fig:marker}c),
including an oil-objective (Olympus, UPLANSAPO, 60$\times$, 1.35~NA),
single photon counting detectors and a 585~nm long-pass filter.
The sample is mounted on a piezo scanner (PI, P-517.3CD with 1~nm in-plane
and 0.1~nm vertical resolution). The excitation power was 0.5~mW onto the
diffraction limited spot (~600~nm). The imaging
depth corresponds to the desired SIL radius, typically 2-6$\mu$m in the present
case. Single emitters were identified by
measuring the autocorrelation function in a Hanbury-Brown and Twiss
configuration. After the emitters were located, the surface of the sample was
imaged to locate the position of the three holes around the NV
accurately.
The accuracy of fabricated SIL is limited by the accuracy of locating markers in
both experimental configurations.
Since the optical signal from a single marker
originates from a sub-wavelength structure, the accuracy for
locating an ideal point source $\sigma_r$ is in theory on the order of
$\sigma_r \sim 0.61\lambda / (N.A.\sqrt{N})$, where $N$ is the number of
detected photons. This amounts to $\sim 1{\rm nm}/\sqrt{\rm Hz}$ for a single
emitter with a detected count rate of 100 kcounts/s.
For non-ideal spherical sources, such as the present alignment markers or for
non-axial dipoles such as the single emitters in the present case, the localization
accuracy is lower. The alignment accuracy is further limited by a (possibly
inhomogeneous) fluorescence background and drift as well as imperfect
repeatability of the piezo scanner.
To examine the accuracy of this alignment procedure, we made the following test.
First, we identified a specific emitter, then three markers around it were
selected and all coordinates were recorded. As next step, the
procedure was repeated and the coordinates of the same
markers and emitter were recorded again. Finally, the expected coordinates of the
emitter were calculated from the previous coodinates of the emitter and the
previous and new coordinates of the markers. This procedure is identical to the
calculation of the target SIL position from the ion beam image. The difference
between the calculated and the measured new position of the emitter defines
roughly the positioning accuracy which was typically better than 100~nm.
In fact, by minimizing mechanical drift and fitting the experimental
data with a Gaussian function it should be possible to achieve positioning
accuracy of several nanometers \cite{Bobroff1986,Thompson2002}. As there was no need for such a
accuracy in our experiment, the positions of the markers were extracted directly
from the brightest pixels in the image.

The lateral coordinates $r'_e$ of the emitter in the ion image follow from the
coordinates of the emitter $r_e$ and the coordinates of the markers $r_0, r_1,
r_2$ in the optical image and the coordinates of the same markers $r'_0, r'_1, r'_2$ in the ion
image as
\begin{equation}
r'_e=r'_0+V'V^{-1}(r_e-r_0),
\end{equation}
where the rows of the $2\times2$ matrices $V$ and
$V'$ are $r_1-r_0$ and $r_2-r_0$, respectively $r'_1-r'_0$ and $r'_2-r'_0$.

\begin{figure}
  \includegraphics[width=70mm]{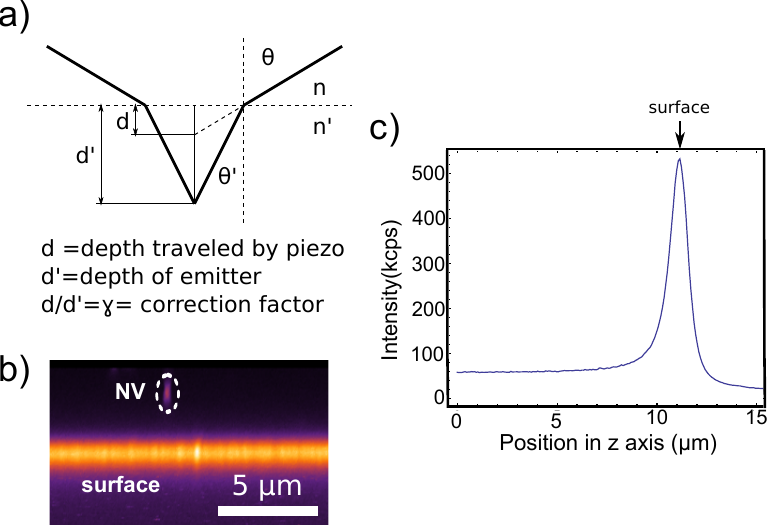}
  \caption{a) Schematic of the focal shift in a high refractive index sample.
  b) Confocal scan of sample in xz plane.
  c) Intensity profile in z direction.}
    \label{fig:surface}
\end{figure}

After discussing the lateral localization of the emitter, we now turn to the
depth determination. Optically, this is not a trivial task, since the
refractive index mismatch between the immersion oil and the diamond
elongates the effective focal length of the microscope objective and
distorts its point spread function. Consider the extremal rays, which are given
by the numerical aperture of the objective lens as shown in
Fig.~\ref{fig:surface}a).
The depth of the emitter $d$ follows from the
displacement $d_0$ of the piezo stage as $d=d_0 \tan
\theta_0/\tan\theta$, where $\theta=\arcsin(\frac{n_0}{n}\sin\theta_0)$ and
$n_0\sin\theta_0=N.A.$, where $n_0$ is the refractive index of the surrounding
medium and $\theta_0$ is the half opening angle.
It is obvious that every ray bundle in the light cone results in a
different focal shift. This is a well known problem in confocal microscopy
studied both experimentally \cite{Carlsson1991,Hell1993} and
theoretically \cite{Visser1992}. For low numerical aperture lenses and small
refractive index mismatch the estimate based on the extremal rays is
reportedly a good approximation \cite{Carlsson1991,Visser1992}.
With high numerical aperture lenses this approximation
should be replaced by a wave optical treatment and the
distorted point spread function should be computed.
The intensity maximum of the point spread function (PSF) would then be
identified with the apparent focus position. Note that in the present case of confocal fluorescence
microscopy of a single emitter a number of additional effects should be taken
into account, including both the illuminating and the emitting field (532 nm
and 650-750 nm, respectively, in the present case), which are additionally
effected by chromatic aberations, the pinhole, and the emission pattern,
determined by two perpendicular dipoles in case of the NV center.
Moreover, in the present case the refractive index
mismatch between the immersion oil ($n_{oil}=1.52$) and
diamond is exceptionally large.
Note that in a full wave model, the effective depth will also no
longer depend linearly on the nominal depth.
Generally, as the PSF gets more and more distorted and the imaging
quality is reduced, the objective lens effectively behaves like a lens
with smaller numerical aperture. Or, equivalently, the extremal rays will
no longer result in constructive interference and contribute less.
We now perform two rough estimates to account for these effects: on the one
hand, we consider the effective depth corresponding to the mean opening angle, given by
$\overline{\theta}=0.5\arcsin(N.A./n)$. On the other hand, we evaluate the
effective depth averaged over all rays that impinge on the back aperture of the
objective with equal weight. The latter corresponds to a simple ray traycing
model. In the present case, we obtain a correction factor $d'/d=\gamma=1.80$ and
$\gamma=2.42$, for the prior and the latter estimate, respectively. We have also checked these results
by full wave optical simulations of the three-dimensional PSF using Zemax,
assuming an ideal objective lens with the same $N.A.$ as the one used in
experiments. In this approach we used the same wavelength (650 nm) for
illumination and emission and disregarded the confocal pinhole. These
simulations confirmed that the conversion factor is depth dependent and in the
range $\gamma\in[2.10,1.95]$ for $d_0\in[0,10]\,\mu$m.
For the experiments $\gamma=1.85$ was used, which led to
good results. While not being subject of the present study one could extend on
this. The effective focus depth could be determined experimentally as follows.
Consider a diamond sample with a thin (few nm) fluorescent layer at a depth several
micrometers below the sample surface (such a sample could be created e.g. by CVD
growth and delta doping\cite{Ohno2012} or by overgrowing a substrate with
a 'dirty' initial surface). Next a staircase structure with well calibrated step
height could be milled into the diamond down to the fluorescent layer by FIB. By
measuring the effective depth of the fluorescent layer on all steps, the depth
dependent effective focus could be reconstructed.

The other crucial factor for the depth detecrmination is to locate the sample
surface precisely. For this purpose we obtained a x-z scan through the sample
(Fig.\ref{fig:surface}b). The confocal image shows a bright band with a narrow
line of higher intensity at its center, also the plot of the z position versus
the intensity of the light shows a sharp peak which is related to the surface
(Fig.\ref{fig:surface}c). Our experimental results prove that our accuracy in
depth determination is better than 1~$\mu$m.

We produce hemispherical SILs with the radius equal to the depth of the NV, and
a cone surrounding the SIL~(Fig.\ref{fig:em}). The cone is chosen slightly
larger than the acceptance cone of the microscope objective, such that the largest possible
amount of light can be captured. More specifically, we use a cone radius
slightly larger than
\begin{equation}
  \dfrac{R_{\rm cone}}{R_{\rm SIL}}=\tan (\arcsin (\dfrac{N.A.}{n'})),
\end{equation}
where $n'$ is the refractive index of the diamond. This ratio is
equal to 2.1 for the oil objective with the N.A. of 1.35.

\section{FIB Milling}
Prior to milling the sample was mounted on a conductive holder with conducting
silver paste and covered with 20~nm conductive gold. Subsequently the
sample is placed inside the FIB machine and aligned by electron microscopy (EM).
Before milling, astigmatism of the ion beam was carefully aligned in a prior
step to optimize the beam shape.

FIB milling was performed on an FEI, Helios 400 machine using a so-called
stream-file input. This
file-format is machine-specific and contains milling times and $x$ and $y$
coordinates.
We wrote scripts in the python programming language to generate stream file
from NV and markers positions, that we include in the
supplementary material. The programs use the width of the FIB image as
a reference to calculate the relative coordinates of the structure. The milling
time for each point is computed based on the given milling rate and the beam
current that is used for the milling.

To define a hemisphere along with a conical cutout, we milled concentric
rings with decreasing inner and outer diameter. Within each ring, we stear the
beam on a double spiral beam path (two interleaved spirals with opposite
handedness; see also supplementary source code) with equidistant points and we
adjust the milling times of the edge points to account for the spherical shape.
Note that a single spiral should yield comparable results.
The number of rings (layers) varies depending on the beam current and size of the SIL.
We ensure that the thickness of each milling layer is much smaller than
the optical wavelength. To ensure homogeneous milling and at the same time
keep the memory usage in the machine within the limits, each layer may
comprise several repetitions of the same path. We also tested milling with
automated drift correction, however, this did not effect on the quality of the
SILs.

With these settings, first the alignment holes are milled as described above.
After this milling, the
sample is sonicated in acetone for 10 min to remove silver paste that was used
for mounting the sample. Then the sample is cleaned in aqua regis to dissolve the
evaporated gold layer. Furthermore the sample was cleaned in piranha solution,
mixture of 1:1 concentrated sulfuric acid and 30 \% hydrogen peroxide solution,
to remove the organic material
from surface and finally, rinsed in deionized water.

\begin{figure}
  \includegraphics[width=70mm]{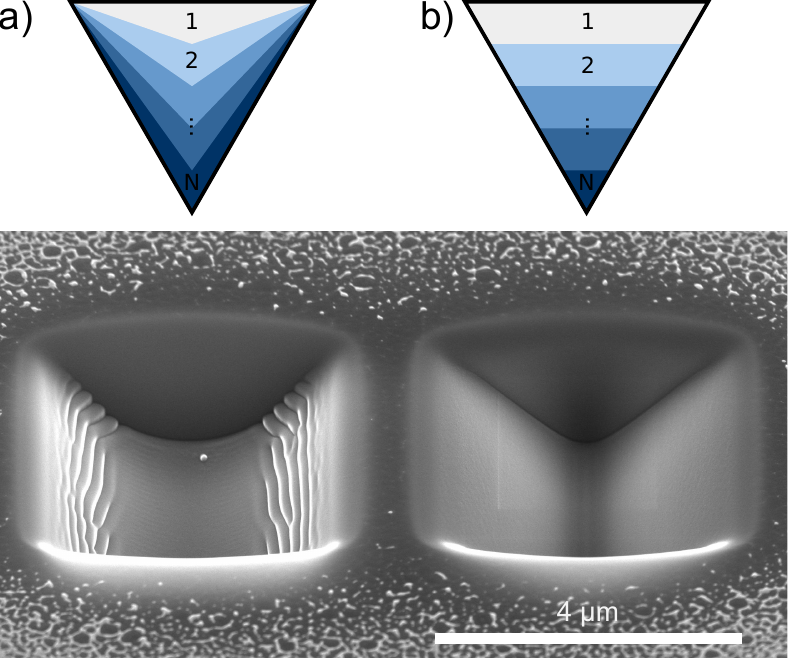}
  \caption{Milling strategy for better focus and less residual. a) Same path, as
shown schematically, was repetitively applied to the sample to mill a structure.
The tip of the structure is round and there are visible residuals in both side
walls. b) Different approach to mill the same structure: as shown schematically
different milling paths were applied layer by layer. The sharpness is increased
and no residuals are visible on the substrate.}
  \label{fig:strategy}
 \end{figure}

There are generally two milling strategies to define a 3D structure as we
illustrate schematically in Fig.~\ref{fig:strategy}). In a first strategy,
$N$ identical milling layers are used. Each milling layer covers the entire
area of the structure and for each milling layer, the depth at each
$x$-$y$-coordinate is $z(x,y)/N$. In a second strategy, the structure is sliced into $N$ layers of
equal thickness $N/z_{max}$ (where $z_{max}$ is the deepest point of the
structure). The layers cover different areas and are milled successively,
starting with the layer that covers the largest area.
In the present case, it was crucial to use the latter strategy. The effect can
be seen with a V-groove as a test structure. As is shown in
Fig.~\ref{fig:strategy}, the first milling strategy results in significant
broadening and milling residuals. By contrast, the second milling
strategy produces a better result. In this case, the rounding of the
dip is roughy given by the beam waist. For all milled SILs the second milling
strategy was used.

\begin{figure}
  \includegraphics[width=70mm]{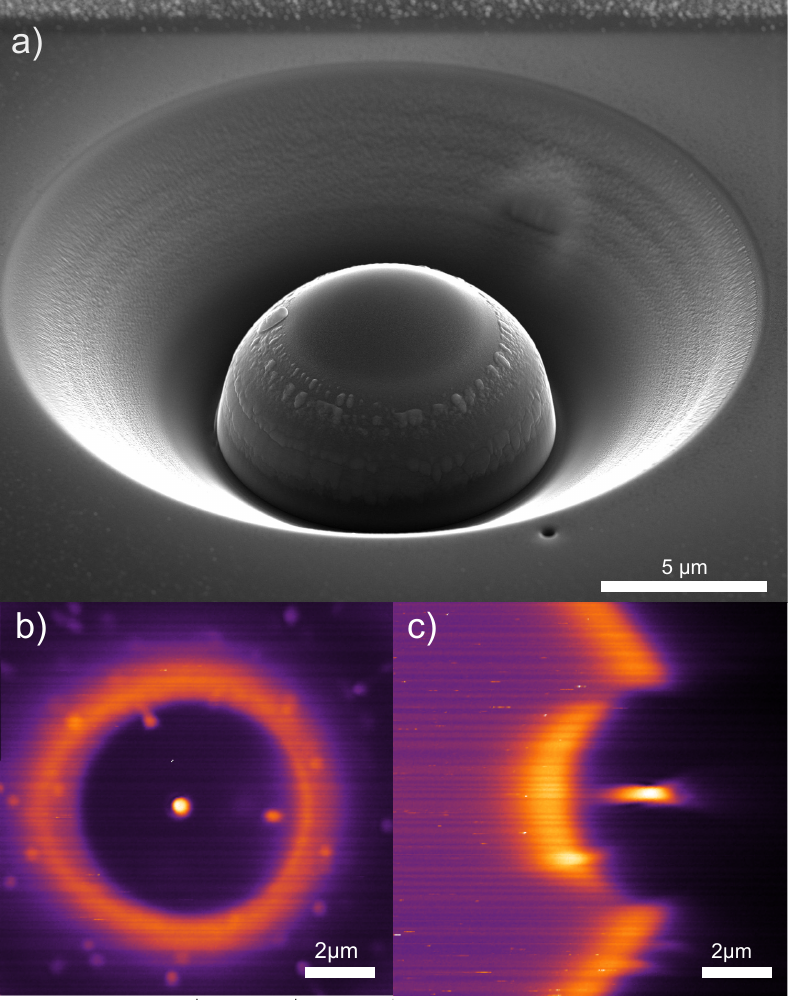}
  \caption{SEM and optical characterization of the milled SIL. a) SEM image under 52$^\circ$ angle.
  b,c) Confocal images of the SIL with the NV center in the lateral and cross
  sectional plane respectively.
  }
  \label{fig:em}
\end{figure}

The markers, which were previously located with confocal microscopy,
were imaged by EM and FIB with lateral resolution of about 30 nm.
After the determination of the marker coordinates in the FIB
image, the relative position of the target NV is calculated in the FIB
coordinates and a stream file for a SIL with radius coinciding with the NV depth is
executed using the calculated lateral NV coordinate as origin.
A typical set of milling parameters for a 10~$\mu$m SIL is outlined in
Table~\ref{table:parameter}.
One of the crucial values is the milling rate. To
calibrate this rate we drilled a cylindrical structure on the sample with the
known diameter, milling time and beam current.
After milling, the depth of the cylinder was measured under 52$^\circ$ angle by
the SEM to calculate the volume of milled area.
Some of the generated stream files have a length of $12.8$ million points to be
cut. The FIB only allows for a maximum number of 8 million points per file. To
overcome this limitation, stream files were automatically split into several files.
The produced files can be loaded into the machine at the same time and ran
automatically after each other. The total milling time for the
given example is 62~minutes and 55~seconds. The SEM image of the milled SIL is
shown in Fig.~\ref{fig:em}a).
\begin{table}
\caption{FIB parameters}
\begin{tabular}{|l|l|}
\hline
Parameter&value for 10~$\mu$m SIL\\ \hline \hline
width of FIB image&64~$\mu m$\\ \hline
milling rate&0.21~$\frac{\mu m^{3}}{\mu s\cdot nA}$\\ \hline
beam current&2.7~nA\\ \hline
radius (SIL)&5~$\mu m$\\ \hline
radius (cone)&2.2$\cdot$radius (SIL)\\ \hline
number of slices&100\\ \hline
\end{tabular}
\label{table:parameter}
\end{table}
After the milling process, the SIL is characterized in the EM. The presented SIL
is a typical example and shows the quality and the overall deviations of the
structure from the ideal shape. The surface roughness is determined to be on the
order of 30~nm peak-to-valley, which was determined by SEM. In early
experiments we milled cross sections through the SIL and carefully checked the
spherical shape. The present stream files do not require any further corrections
regarding the spherical shape of the SILs.
Before any further optical characterization, the cleaning processes as explained
before was applied. Additionally, the sample was boiled 3 hours in the mixture
of concentrated sulphuric acid, nitric acid and perchloric acid in a volume ratio
1:1:1 to remove the residual materials and implanted gallium. While this
cleaning procedure removes all background fluorescence, likely there is
still implanted gallium present within the first 10 nanometers underneath the
diamond surface \cite{widmann_coherent_2014}.
\section{Optical Characterization}
\begin{figure}
  \includegraphics[width=70mm]{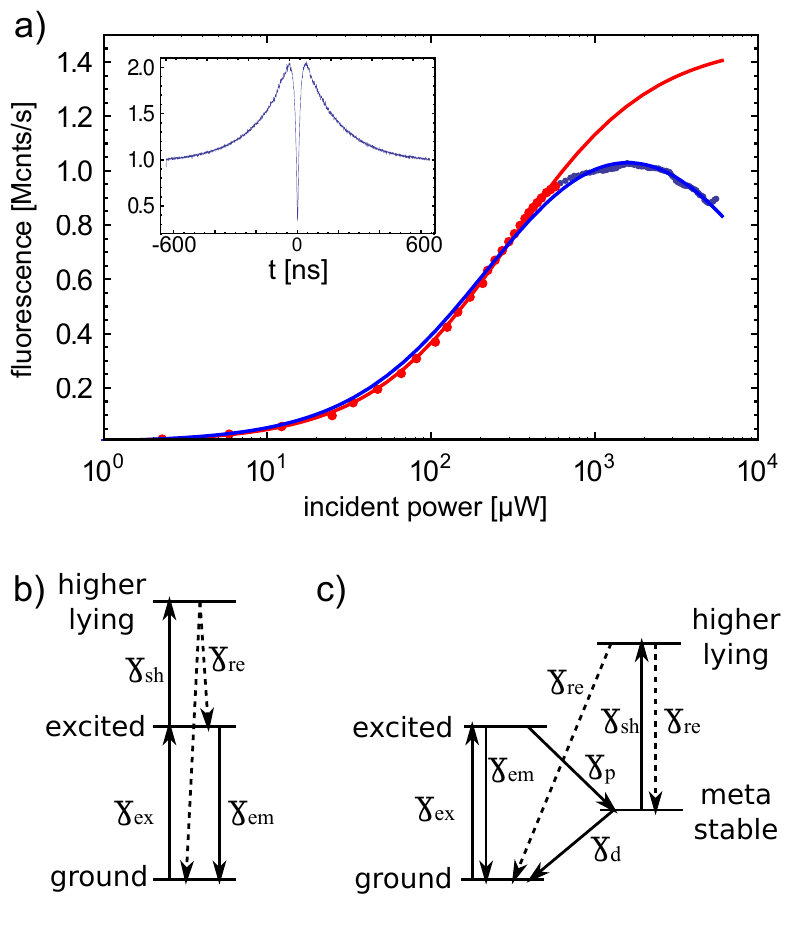}
  \caption{Saturation behaviour. a) The saturation curve of NV in the
  center of the SIL (red and blue dots), The red dots show the data up to
  $550\,\mu$W, the blue dots show the full data set.
  The red and blue line show fits with a two-level model and a three-level model
  with higher lying excited state, respectively (see text for details).
  The inserted plot shows the antibunching at high laser power. b),c)
  Three- and four-level rate equation models with spontaneously decaying higher
  lying excited states.}
  \label{fig:saturation}
\end{figure}

The resulting SILs were characterized optically in a confocal fluorescence
microscope. For our studies, we used a [111] cut diamond and a single NV center
oriented perpendicular to the diamond surface. In this case, the
two optical dipoles are oriented parallel to the surface, resulting in
an optimal directivity of the dipolar emission pattern.
Since we have studied all emitters prior to FIB-milling, it is possible to
compare the results prior and after the milling process. The central aim is to
optimize the detectable count-rates.
Optical studies after milling are presented in Fig.~\ref{fig:em}b) and c). The
deviation of the emitter from the center of the SIL is less than 150~nm.
The PSF is found to be circular. In the linear excitation regime, the
signal to background ratio is larger than 30. At higher laser powers, the background
dominates, since it rises linearly with excitation power, while the single
emitter is saturated. A saturated count rate of 1.0 Mcounts/s was observed,
utilizing an oil objective~(Fig.~\ref{fig:saturation}a), which is the highest
count rate reported for a single NV in a bulk diamond so far. Compared to the
saturated count rate without the SIL, 350 kcounts/s (data not shown), this is an
enhancement $\xi=3.4$. This agrees well with the expected enhancement
\begin{equation}
\xi=\frac{1-\cos(\arcsin(N.A./n))}{1-\cos(\arcsin(N.A./n'))}=3.3.
\end{equation}
Here $n$ and $n'$ are the refractive indices of the oil and diamond,
respectively. The optical
characterization was repeated with an air objective (Nikon, CFI LU Plan Fluor
EPI P 100$\times$, 0.9~NA), were we observed count rates of ~65 kcnts/s and ~600
kcnts/s, without and with the SIL, respectively.
Based on the specified numerical apertures,
the collection angle of the oil and air objectives should be the same.
This substantial difference in count rate cannot be explained by the increase of
the reflection losses at the air-diamond interface, which is less than 10\%.
A possible explanation for this observation could be slight deviations of the
SIL from the ideal hemispherical shape that are less critical when using an oil
objective. A second explanation could be that there are substantial differences
of the optical wave fronts, even though the manufacturers specify equivalent
collection angles.

The acquired antibunching curve \cite{kurtsiefer_rpl_2000} (see inset in
Fig.~\ref{fig:saturation}a) clearly shows that we have a single emitter.
The presented $g^{(2)}(\tau)$-function was background corrected as
outlined in literature~\cite{kitson_pra_1998}. It further allows to
estimate the decay rates of the excited state and metastable triplet state of
the NV center. We find the following rates:
$T_1=$ 9~ns, $T_m=$ 236~ns, which is in good agreement with previous studies.
To quantify the achievable count rate, we measure a saturation curve
shown Fig.~\ref{fig:saturation}a). The low intensity part of the data is well
described by
\begin{equation}
I_{em}=I_{\inf}\frac{I_0}{I_0+I_{sat}},
\end{equation}
where
$I_{em}$ is the emitted intensity, $I_0$ is the incident intensity,
$I_{sat}$ is the saturation intensity and $I_{\inf}$ is the saturated
fluorescence intensity.
The above model commonly describes a two-level system, a three-level system with a
long lived metastable singlet level, a five-level system accounting for the
different spin states of the NV center, as well as a simple four-level system
accounting for optically induced charge state switching between NV$^0$ and
NV$^-$\cite{Beha2012}.
With realistic parameters, all of these models result in qualitatively the same
saturation curve. However, at high excitation powers, these models fail to
describe the present data. Indeed, in our data we observe that the fluorescence
decreases towards high incident powers. Such behaviour has
been reported previously \cite{siyushev_apl_2010,Han2012} and was attributed
to the existence of dark states that could be described by a six-level system.
Here we propose two simpler models that allow us to describe the
saturation behavior at high incident powers. Note that similar models have been
exploited to explain the population dynamics of molecules,
chromium~\cite{Aharonovich2010} and SiV~\cite{Neu2012} centers in diamond.
The models are shown schematically in
Fig.\ref{fig:saturation} b) and c). We assume that there exists a higher lying
excited state that is populated optically and
decays spontaneously either back to the same state or to the ground state,
indicated by the dashed arrows (within the limits considered below, both cases
result in the same saturation behaviour). The higher lying excited statee
behaves like a shelving state.
Note that it is key that the decay of the higher lying state is
spontaneous. This is in contrast to the NV$^-$ to NV$^0$ switching mechanism,
where both the ionization and the recovery path are believed to be driven by optical
pumping~\cite{Beha2012}. The higher lying state could
be coupled either to the excited state or to the metastable singlet state.
Thus, the first model employs a
three-level system with ground and excited state and an additional
higher lying excited state that can be populated optically from the lower
excited state and decays spontaneously (the meta stable state is
omitted here, to provide the simplest possible model).
The second model employs a four-level system with ground, excited and
metastable singlet state and a higher lying
excited state that can be populated from the singlet state and decays
spontaneously back to the singlet state. The models are characterized by the
transition rates between the states, which are the excitation and emission
rates between the
ground and excited state, the excitation and recovery rates to and from the
higher lying state and the population and decay rate of metastable state. The
rates are denoted $\gamma_{ex}$, $\gamma_{em}$, $\gamma_{sh}$,
$\gamma_{re}$, $\gamma_{p}$, and $\gamma_{d}$, respectively. The excitation
rates are proportional to the incident intensity and the
corresponding efficiencies: $\gamma_{ex}(I_0)=\eta_{ex}I_0$ and
$\gamma_{sh}(I_0)=\eta_{sh}I_0$. The corresponding rate equations result in
saturation curves
\begin{equation}
I_{em}\propto\frac{\gamma_{re}\eta_{ex}I_0}{\gamma_{em}\gamma_{re}+\gamma_{re}\eta_{ex}I_0+\eta_{ex}\eta_{sh}I_0^2}
\end{equation}
and
\begin{equation}
I_{em}\propto\frac{\gamma_{d}\gamma_{re}\eta_{ex}I_0}{(\gamma_{p}+\gamma_{em})\gamma_{d}\gamma_{re}+(\gamma_{p}+\gamma_{d})\gamma_{re}\eta_{ex}I_0+\gamma_{p}\eta_{ex}\eta_{sh}I_0^2}
\end{equation}
for the three- and four-level model, respectively. Both models
result in the same expression for the saturation curve with three free
constants. We fit the experimental data with this common expression. The result
is shown by the blue curve in Fig.\ref{fig:saturation}.

We conjecture that quite generally, the introduction of a higher lying excited
state that is populated by optical pumping but that decays spontaneously will
result in the observed saturation behavior.

\section{Conclusion}
In this paper, we describe our procedure to precisely locate single emitters
both optically inside a diamond substrate and in a focused ion beam machine.
FIB milled holes have proven suitable alignment markers, providing $100$ nm
lateral and $500$ nm axial accuracy.
We have compared different FIB milling strategies. Milling layers of equal
thickness yielded the best result. This strategy might be applied to other
hard materials, to ensure a minimal amount of residuals and optimal feature
sharpness. We present record-high count
rates of NV at the focus of a SIL.
The technique paves the way for high-brightness single photon sources based on solid
state emitters and novel spin-control schemes.

\begin{acknowledgments}
We thank Florian Dolde and Samuel Wagner for help with spin measurements in
the SILs and Sen Yang, Thai-Hien Tran and Andrej Denisenko for fruitful discussions.
The hospitality of the ITO, Stuttgart is acknowledged. We acknowledge financial
support by the Max Planck Society, the ERC project SQUTEC, the DFG SFB/TR21,
the EU projects DIAMANT, SIQS, QESSENCE and QINVC, the JST-DFG (FOR1482 and
FOR1493), the Volkswagenstiftung and the Landesstiftung Baden-w{\"u}rttemberg.
\end{acknowledgments}

\bibliography{sil}

\begin{thebibliography}{32}%
\makeatletter
\providecommand \@ifxundefined [1]{%
 \@ifx{#1\undefined}
}%
\providecommand \@ifnum [1]{%
 \ifnum #1\expandafter \@firstoftwo
 \else \expandafter \@secondoftwo
 \fi
}%
\providecommand \@ifx [1]{%
 \ifx #1\expandafter \@firstoftwo
 \else \expandafter \@secondoftwo
 \fi
}%
\providecommand \natexlab [1]{#1}%
\providecommand \enquote  [1]{``#1''}%
\providecommand \bibnamefont  [1]{#1}%
\providecommand \bibfnamefont [1]{#1}%
\providecommand \citenamefont [1]{#1}%
\providecommand \href@noop [0]{\@secondoftwo}%
\providecommand \href [0]{\begingroup \@sanitize@url \@href}%
\providecommand \@href[1]{\@@startlink{#1}\@@href}%
\providecommand \@@href[1]{\endgroup#1\@@endlink}%
\providecommand \@sanitize@url [0]{\catcode `\\12\catcode `\$12\catcode
  `\&12\catcode `\#12\catcode `\^12\catcode `\_12\catcode `\%12\relax}%
\providecommand \@@startlink[1]{}%
\providecommand \@@endlink[0]{}%
\providecommand \url  [0]{\begingroup\@sanitize@url \@url }%
\providecommand \@url [1]{\endgroup\@href {#1}{\urlprefix }}%
\providecommand \urlprefix  [0]{URL }%
\providecommand \Eprint [0]{\href }%
\providecommand \doibase [0]{http://dx.doi.org/}%
\providecommand \selectlanguage [0]{\@gobble}%
\providecommand \bibinfo  [0]{\@secondoftwo}%
\providecommand \bibfield  [0]{\@secondoftwo}%
\providecommand \translation [1]{[#1]}%
\providecommand \BibitemOpen [0]{}%
\providecommand \bibitemStop [0]{}%
\providecommand \bibitemNoStop [0]{.\EOS\space}%
\providecommand \EOS [0]{\spacefactor3000\relax}%
\providecommand \BibitemShut  [1]{\csname bibitem#1\endcsname}%
\let\auto@bib@innerbib\@empty
\bibitem [{\citenamefont {Moerner}\ and\ \citenamefont
  {Kador}(1989)}]{moerner_prl_1989}%
  \BibitemOpen
  \bibfield  {author} {\bibinfo {author} {\bibfnamefont {W.~E.}\ \bibnamefont
  {Moerner}}\ and\ \bibinfo {author} {\bibfnamefont {L.}~\bibnamefont
  {Kador}},\ }\href {\doibase 10.1103/PhysRevLett.62.2535} {\bibfield
  {journal} {\bibinfo  {journal} {Phys. Rev. Lett.}\ }\textbf {\bibinfo
  {volume} {62}},\ \bibinfo {pages} {2535} (\bibinfo {year}
  {1989})}\BibitemShut {NoStop}%
\bibitem [{\citenamefont {Jelezko}\ and\ \citenamefont
  {Wrachtrup}(2006)}]{jelezko_pssa_2006}%
  \BibitemOpen
  \bibfield  {author} {\bibinfo {author} {\bibfnamefont {F.}~\bibnamefont
  {Jelezko}}\ and\ \bibinfo {author} {\bibfnamefont {J.}~\bibnamefont
  {Wrachtrup}},\ }\href {\doibase 10.1002/pssa.200671403} {\bibfield  {journal}
  {\bibinfo  {journal} {physica status solidi a}\ }\textbf {\bibinfo {volume}
  {203}},\ \bibinfo {pages} {3207} (\bibinfo {year} {2006})}\BibitemShut
  {NoStop}%
\bibitem [{\citenamefont {Dutt}\ \emph {et~al.}(2007)\citenamefont {Dutt},
  \citenamefont {Childress}, \citenamefont {Jiang}, \citenamefont {Togan},
  \citenamefont {Maze}, \citenamefont {Jelezko}, \citenamefont {Zibrov},
  \citenamefont {Hemmer},\ and\ \citenamefont {Lukin}}]{dutt_science_2007}%
  \BibitemOpen
  \bibfield  {author} {\bibinfo {author} {\bibfnamefont {M.~V.~G.}\
  \bibnamefont {Dutt}}, \bibinfo {author} {\bibfnamefont {L.}~\bibnamefont
  {Childress}}, \bibinfo {author} {\bibfnamefont {L.}~\bibnamefont {Jiang}},
  \bibinfo {author} {\bibfnamefont {E.}~\bibnamefont {Togan}}, \bibinfo
  {author} {\bibfnamefont {J.}~\bibnamefont {Maze}}, \bibinfo {author}
  {\bibfnamefont {F.}~\bibnamefont {Jelezko}}, \bibinfo {author} {\bibfnamefont
  {A.~S.}\ \bibnamefont {Zibrov}}, \bibinfo {author} {\bibfnamefont {P.~R.}\
  \bibnamefont {Hemmer}}, \ and\ \bibinfo {author} {\bibfnamefont {M.~D.}\
  \bibnamefont {Lukin}},\ }\href@noop {} {\bibfield  {journal} {\bibinfo
  {journal} {Science}\ }\textbf {\bibinfo {volume} {316}},\ \bibinfo {pages}
  {1312} (\bibinfo {year} {2007})}\BibitemShut {NoStop}%
\bibitem [{\citenamefont {Brouri}\ \emph {et~al.}(2000)\citenamefont {Brouri},
  \citenamefont {Beveratos}, \citenamefont {Poizat},\ and\ \citenamefont
  {Grangier}}]{Brouri_opt_2000}%
  \BibitemOpen
  \bibfield  {author} {\bibinfo {author} {\bibfnamefont {R.}~\bibnamefont
  {Brouri}}, \bibinfo {author} {\bibfnamefont {A.}~\bibnamefont {Beveratos}},
  \bibinfo {author} {\bibfnamefont {J.-P.}\ \bibnamefont {Poizat}}, \ and\
  \bibinfo {author} {\bibfnamefont {P.}~\bibnamefont {Grangier}},\ }\href
  {\doibase 10.1364/OL.25.001294} {\bibfield  {journal} {\bibinfo  {journal}
  {Opt. Lett.}\ }\textbf {\bibinfo {volume} {25}},\ \bibinfo {pages} {1294}
  (\bibinfo {year} {2000})}\BibitemShut {NoStop}%
\bibitem [{\citenamefont {Kurtsiefer}\ \emph {et~al.}(2000)\citenamefont
  {Kurtsiefer}, \citenamefont {Mayer}, \citenamefont {Zarda},\ and\
  \citenamefont {Weinfurter}}]{kurtsiefer_rpl_2000}%
  \BibitemOpen
  \bibfield  {author} {\bibinfo {author} {\bibfnamefont {C.}~\bibnamefont
  {Kurtsiefer}}, \bibinfo {author} {\bibfnamefont {S.}~\bibnamefont {Mayer}},
  \bibinfo {author} {\bibfnamefont {P.}~\bibnamefont {Zarda}}, \ and\ \bibinfo
  {author} {\bibfnamefont {H.}~\bibnamefont {Weinfurter}},\ }\href@noop {}
  {\bibfield  {journal} {\bibinfo  {journal} {Phys. Rev. Lett.}\ }\textbf
  {\bibinfo {volume} {85}},\ \bibinfo {pages} {290} (\bibinfo {year}
  {2000})}\BibitemShut {NoStop}%
\bibitem [{\citenamefont {Balasubramanian}\ \emph {et~al.}(2008)\citenamefont
  {Balasubramanian}, \citenamefont {Chan}, \citenamefont {Kolesov},
  \citenamefont {Al-Hmoud}, \citenamefont {Tisler}, \citenamefont {Shin},
  \citenamefont {Kim}, \citenamefont {Wojcik}, \citenamefont {Hemmer},
  \citenamefont {Krueger}, \citenamefont {Hanke}, \citenamefont
  {Leitenstorfer}, \citenamefont {Bratschitsch}, \citenamefont {Jelezko},\ and\
  \citenamefont {Wrachtrup}}]{balasubramanian_nature_2008}%
  \BibitemOpen
  \bibfield  {author} {\bibinfo {author} {\bibfnamefont {G.}~\bibnamefont
  {Balasubramanian}}, \bibinfo {author} {\bibfnamefont {I.~Y.}\ \bibnamefont
  {Chan}}, \bibinfo {author} {\bibfnamefont {R.}~\bibnamefont {Kolesov}},
  \bibinfo {author} {\bibfnamefont {M.}~\bibnamefont {Al-Hmoud}}, \bibinfo
  {author} {\bibfnamefont {J.}~\bibnamefont {Tisler}}, \bibinfo {author}
  {\bibfnamefont {C.}~\bibnamefont {Shin}}, \bibinfo {author} {\bibfnamefont
  {C.}~\bibnamefont {Kim}}, \bibinfo {author} {\bibfnamefont {A.}~\bibnamefont
  {Wojcik}}, \bibinfo {author} {\bibfnamefont {P.~R.}\ \bibnamefont {Hemmer}},
  \bibinfo {author} {\bibfnamefont {A.}~\bibnamefont {Krueger}}, \bibinfo
  {author} {\bibfnamefont {T.}~\bibnamefont {Hanke}}, \bibinfo {author}
  {\bibfnamefont {A.}~\bibnamefont {Leitenstorfer}}, \bibinfo {author}
  {\bibfnamefont {R.}~\bibnamefont {Bratschitsch}}, \bibinfo {author}
  {\bibfnamefont {F.}~\bibnamefont {Jelezko}}, \ and\ \bibinfo {author}
  {\bibfnamefont {J.}~\bibnamefont {Wrachtrup}},\ }\href@noop {} {\bibfield
  {journal} {\bibinfo  {journal} {Nature}\ }\textbf {\bibinfo {volume} {455}},\
  \bibinfo {pages} {648} (\bibinfo {year} {2008})}\BibitemShut {NoStop}%
\bibitem [{\citenamefont {Koyama}\ \emph {et~al.}(1999)\citenamefont {Koyama},
  \citenamefont {Yoshita}, \citenamefont {Baba}, \citenamefont {Suemoto},\ and\
  \citenamefont {Akiyama}}]{koyama_apl_1999}%
  \BibitemOpen
  \bibfield  {author} {\bibinfo {author} {\bibfnamefont {K.}~\bibnamefont
  {Koyama}}, \bibinfo {author} {\bibfnamefont {M.}~\bibnamefont {Yoshita}},
  \bibinfo {author} {\bibfnamefont {M.}~\bibnamefont {Baba}}, \bibinfo {author}
  {\bibfnamefont {T.}~\bibnamefont {Suemoto}}, \ and\ \bibinfo {author}
  {\bibfnamefont {H.}~\bibnamefont {Akiyama}},\ }\href {\doibase
  10.1063/1.124833} {\bibfield  {journal} {\bibinfo  {journal} {Applied Physics
  Letters}\ }\textbf {\bibinfo {volume} {75}},\ \bibinfo {pages} {1667}
  (\bibinfo {year} {1999})}\BibitemShut {NoStop}%
\bibitem [{\citenamefont {Lee}\ \emph {et~al.}(2011)\citenamefont {Lee},
  \citenamefont {Chen}, \citenamefont {H.}, \citenamefont {P.}, \citenamefont
  {Lettow}, \citenamefont {Renn}, \citenamefont {Sandoghdar},\ and\
  \citenamefont {Gotzinger}}]{lee_np_2011}%
  \BibitemOpen
  \bibfield  {author} {\bibinfo {author} {\bibfnamefont {K.}~\bibnamefont
  {Lee}}, \bibinfo {author} {\bibfnamefont {X.}~\bibnamefont {Chen}}, \bibinfo
  {author} {\bibfnamefont {E.}~\bibnamefont {H.}}, \bibinfo {author}
  {\bibfnamefont {K.}~\bibnamefont {P.}}, \bibinfo {author} {\bibfnamefont
  {R.}~\bibnamefont {Lettow}}, \bibinfo {author} {\bibfnamefont
  {A.}~\bibnamefont {Renn}}, \bibinfo {author} {\bibfnamefont {V.}~\bibnamefont
  {Sandoghdar}}, \ and\ \bibinfo {author} {\bibfnamefont {S.}~\bibnamefont
  {Gotzinger}},\ }\href {http://dx.doi.org/10.1038/nphoton.2010.312} {\bibfield
   {journal} {\bibinfo  {journal} {Nat Photon}\ }\textbf {\bibinfo {volume}
  {5}},\ \bibinfo {pages} {166} (\bibinfo {year} {2011})}\BibitemShut {NoStop}%
\bibitem [{\citenamefont {Barnes}\ \emph {et~al.}(2002)\citenamefont {Barnes},
  \citenamefont {Bj{\"o}rk}, \citenamefont {G{\'e}rard}, \citenamefont
  {Jonsson}, \citenamefont {Wasey}, \citenamefont {Worthing},\ and\
  \citenamefont {Zwiller}}]{barnes_epjd_2002}%
  \BibitemOpen
  \bibfield  {author} {\bibinfo {author} {\bibfnamefont {W.}~\bibnamefont
  {Barnes}}, \bibinfo {author} {\bibfnamefont {G.}~\bibnamefont {Bj{\"o}rk}},
  \bibinfo {author} {\bibfnamefont {J.}~\bibnamefont {G{\'e}rard}}, \bibinfo
  {author} {\bibfnamefont {P.}~\bibnamefont {Jonsson}}, \bibinfo {author}
  {\bibfnamefont {J.}~\bibnamefont {Wasey}}, \bibinfo {author} {\bibfnamefont
  {P.}~\bibnamefont {Worthing}}, \ and\ \bibinfo {author} {\bibfnamefont
  {V.}~\bibnamefont {Zwiller}},\ }\href
  {http://dx.doi.org/10.1140/epjd/e20020024} {\bibfield  {journal} {\bibinfo
  {journal} {Eur. Phys. J. D}\ }\textbf {\bibinfo {volume} {18}},\ \bibinfo
  {pages} {197} (\bibinfo {year} {2002})}\BibitemShut {NoStop}%
\bibitem [{\citenamefont {Babinec}\ \emph {et~al.}(2010)\citenamefont
  {Babinec}, \citenamefont {M.}, \citenamefont {Khan}, \citenamefont {Zhang},
  \citenamefont {Maze}, \citenamefont {Hemmer},\ and\ \citenamefont
  {Loncar}}]{babinec_nn_2010}%
  \BibitemOpen
  \bibfield  {author} {\bibinfo {author} {\bibfnamefont {T.~M.}\ \bibnamefont
  {Babinec}}, \bibinfo {author} {\bibfnamefont {H.~J.}\ \bibnamefont {M.}},
  \bibinfo {author} {\bibfnamefont {M.}~\bibnamefont {Khan}}, \bibinfo {author}
  {\bibfnamefont {Y.}~\bibnamefont {Zhang}}, \bibinfo {author} {\bibfnamefont
  {J.~R.}\ \bibnamefont {Maze}}, \bibinfo {author} {\bibfnamefont {P.~R.}\
  \bibnamefont {Hemmer}}, \ and\ \bibinfo {author} {\bibfnamefont
  {M.}~\bibnamefont {Loncar}},\ }\href {http://dx.doi.org/10.1038/nnano.2010.6}
  {\bibfield  {journal} {\bibinfo  {journal} {Nat Nano}\ }\textbf {\bibinfo
  {volume} {5}},\ \bibinfo {pages} {195} (\bibinfo {year} {2010})}\BibitemShut
  {NoStop}%
\bibitem [{\citenamefont {Mansfield}\ and\ \citenamefont
  {Kino}(1990)}]{mansfield_apl_1990}%
  \BibitemOpen
  \bibfield  {author} {\bibinfo {author} {\bibfnamefont {S.~M.}\ \bibnamefont
  {Mansfield}}\ and\ \bibinfo {author} {\bibfnamefont {G.~S.}\ \bibnamefont
  {Kino}},\ }\href {\doibase 10.1063/1.103828} {\bibfield  {journal} {\bibinfo
  {journal} {Applied Physics Letters}\ }\textbf {\bibinfo {volume} {57}},\
  \bibinfo {pages} {2615} (\bibinfo {year} {1990})}\BibitemShut {NoStop}%
\bibitem [{\citenamefont {Wrigge}\ \emph {et~al.}(2008)\citenamefont {Wrigge},
  \citenamefont {Gerhardt}, \citenamefont {Hwang}, \citenamefont {Zumofen},\
  and\ \citenamefont {Sandoghdar}}]{wrigge_np_2008}%
  \BibitemOpen
  \bibfield  {author} {\bibinfo {author} {\bibfnamefont {G.}~\bibnamefont
  {Wrigge}}, \bibinfo {author} {\bibfnamefont {I.}~\bibnamefont {Gerhardt}},
  \bibinfo {author} {\bibfnamefont {J.}~\bibnamefont {Hwang}}, \bibinfo
  {author} {\bibfnamefont {G.}~\bibnamefont {Zumofen}}, \ and\ \bibinfo
  {author} {\bibfnamefont {V.}~\bibnamefont {Sandoghdar}},\ }\href
  {http://dx.doi.org/10.1038/nphys812} {\bibfield  {journal} {\bibinfo
  {journal} {Nat Phys}\ }\textbf {\bibinfo {volume} {4}},\ \bibinfo {pages}
  {60} (\bibinfo {year} {2008})}\BibitemShut {NoStop}%
\bibitem [{\citenamefont {Vamivakas}\ \emph {et~al.}(2007)\citenamefont
  {Vamivakas}, \citenamefont {Atatuere}, \citenamefont {Dreiser}, \citenamefont
  {Yilmaz}, \citenamefont {Badolato}, \citenamefont {Swan}, \citenamefont
  {Goldberg}, \citenamefont {Imamoglu},\ and\ \citenamefont
  {Unlu}}]{vamivakas_nl_2007}%
  \BibitemOpen
  \bibfield  {author} {\bibinfo {author} {\bibfnamefont {A.~N.}\ \bibnamefont
  {Vamivakas}}, \bibinfo {author} {\bibfnamefont {M.}~\bibnamefont {Atatuere}},
  \bibinfo {author} {\bibfnamefont {J.}~\bibnamefont {Dreiser}}, \bibinfo
  {author} {\bibfnamefont {S.~T.}\ \bibnamefont {Yilmaz}}, \bibinfo {author}
  {\bibfnamefont {A.}~\bibnamefont {Badolato}}, \bibinfo {author}
  {\bibfnamefont {A.~K.}\ \bibnamefont {Swan}}, \bibinfo {author}
  {\bibfnamefont {B.~B.}\ \bibnamefont {Goldberg}}, \bibinfo {author}
  {\bibfnamefont {A.}~\bibnamefont {Imamoglu}}, \ and\ \bibinfo {author}
  {\bibfnamefont {M.~S.}\ \bibnamefont {Unlu}},\ }\href {\doibase
  10.1021/nl0717255} {\bibfield  {journal} {\bibinfo  {journal} {Nano Letters}\
  }\textbf {\bibinfo {volume} {7}},\ \bibinfo {pages} {2892} (\bibinfo {year}
  {2007})},\ \Eprint
  {http://arxiv.org/abs/http://pubs.acs.org/doi/pdf/10.1021/nl0717255}
  {http://pubs.acs.org/doi/pdf/10.1021/nl0717255} \BibitemShut {NoStop}%
\bibitem [{\citenamefont {Robledo}\ \emph {et~al.}(2011)\citenamefont
  {Robledo}, \citenamefont {Childress}, \citenamefont {Bernien}, \citenamefont
  {Hensen}, \citenamefont {Alkemade},\ and\ \citenamefont
  {Hanson}}]{robledo_n_2011}%
  \BibitemOpen
  \bibfield  {author} {\bibinfo {author} {\bibfnamefont {L.}~\bibnamefont
  {Robledo}}, \bibinfo {author} {\bibfnamefont {L.}~\bibnamefont {Childress}},
  \bibinfo {author} {\bibfnamefont {H.}~\bibnamefont {Bernien}}, \bibinfo
  {author} {\bibfnamefont {B.}~\bibnamefont {Hensen}}, \bibinfo {author}
  {\bibfnamefont {P.~F.~A.}\ \bibnamefont {Alkemade}}, \ and\ \bibinfo {author}
  {\bibfnamefont {R.}~\bibnamefont {Hanson}},\ }\href
  {http://dx.doi.org/10.1038/nature10401} {\bibfield  {journal} {\bibinfo
  {journal} {Nature}\ }\textbf {\bibinfo {volume} {477}},\ \bibinfo {pages}
  {574} (\bibinfo {year} {2011})}\BibitemShut {NoStop}%
\bibitem [{\citenamefont {Waldherr}\ \emph {et~al.}(2014)\citenamefont
  {Waldherr}, \citenamefont {Wang}, \citenamefont {Zaiser}, \citenamefont
  {Jamali}, \citenamefont {Schulte-Herbr\"{u}ggen}, \citenamefont {Abe},
  \citenamefont {Ohshima}, \citenamefont {Isoya}, \citenamefont {Du},
  \citenamefont {Neumann},\ and\ \citenamefont {Wrachtrup}}]{Waldherr2014}%
  \BibitemOpen
  \bibfield  {author} {\bibinfo {author} {\bibfnamefont {G.}~\bibnamefont
  {Waldherr}}, \bibinfo {author} {\bibfnamefont {Y.}~\bibnamefont {Wang}},
  \bibinfo {author} {\bibfnamefont {S.}~\bibnamefont {Zaiser}}, \bibinfo
  {author} {\bibfnamefont {M.}~\bibnamefont {Jamali}}, \bibinfo {author}
  {\bibfnamefont {T.}~\bibnamefont {Schulte-Herbr\"{u}ggen}}, \bibinfo {author}
  {\bibfnamefont {H.}~\bibnamefont {Abe}}, \bibinfo {author} {\bibfnamefont
  {T.}~\bibnamefont {Ohshima}}, \bibinfo {author} {\bibfnamefont
  {J.}~\bibnamefont {Isoya}}, \bibinfo {author} {\bibfnamefont {J.~F.}\
  \bibnamefont {Du}}, \bibinfo {author} {\bibfnamefont {P.}~\bibnamefont
  {Neumann}}, \ and\ \bibinfo {author} {\bibfnamefont {J.}~\bibnamefont
  {Wrachtrup}},\ }\href {\doibase 10.1038/nature12919} {\bibfield  {journal}
  {\bibinfo  {journal} {Nature}\ }\textbf {\bibinfo {volume} {506}},\ \bibinfo
  {pages} {204} (\bibinfo {year} {2014})}\BibitemShut {NoStop}%
\bibitem [{\citenamefont {Kolesov}\ \emph {et~al.}(2013)\citenamefont
  {Kolesov}, \citenamefont {Xia}, \citenamefont {Reuter}, \citenamefont
  {Jamali}, \citenamefont {St\"ohr}, \citenamefont {Inal}, \citenamefont
  {Siyushev},\ and\ \citenamefont {Wrachtrup}}]{Kolesov2013}%
  \BibitemOpen
  \bibfield  {author} {\bibinfo {author} {\bibfnamefont {R.}~\bibnamefont
  {Kolesov}}, \bibinfo {author} {\bibfnamefont {K.}~\bibnamefont {Xia}},
  \bibinfo {author} {\bibfnamefont {R.}~\bibnamefont {Reuter}}, \bibinfo
  {author} {\bibfnamefont {M.}~\bibnamefont {Jamali}}, \bibinfo {author}
  {\bibfnamefont {R.}~\bibnamefont {St\"ohr}}, \bibinfo {author} {\bibfnamefont
  {T.}~\bibnamefont {Inal}}, \bibinfo {author} {\bibfnamefont {P.}~\bibnamefont
  {Siyushev}}, \ and\ \bibinfo {author} {\bibfnamefont {J.}~\bibnamefont
  {Wrachtrup}},\ }\href {\doibase 10.1103/PhysRevLett.111.120502} {\bibfield
  {journal} {\bibinfo  {journal} {Phys. Rev. Lett.}\ }\textbf {\bibinfo
  {volume} {111}},\ \bibinfo {pages} {120502} (\bibinfo {year}
  {2013})}\BibitemShut {NoStop}%
\bibitem [{\citenamefont {Naydenov}\ \emph {et~al.}(2011)\citenamefont
  {Naydenov}, \citenamefont {Dolde}, \citenamefont {Hall}, \citenamefont
  {Shin}, \citenamefont {Fedder}, \citenamefont {Hollenberg}, \citenamefont
  {Jelezko},\ and\ \citenamefont {Wrachtrup}}]{Naydenov2011}%
  \BibitemOpen
  \bibfield  {author} {\bibinfo {author} {\bibfnamefont {B.}~\bibnamefont
  {Naydenov}}, \bibinfo {author} {\bibfnamefont {F.}~\bibnamefont {Dolde}},
  \bibinfo {author} {\bibfnamefont {L.~T.}\ \bibnamefont {Hall}}, \bibinfo
  {author} {\bibfnamefont {C.}~\bibnamefont {Shin}}, \bibinfo {author}
  {\bibfnamefont {H.}~\bibnamefont {Fedder}}, \bibinfo {author} {\bibfnamefont
  {L.~C.~L.}\ \bibnamefont {Hollenberg}}, \bibinfo {author} {\bibfnamefont
  {F.}~\bibnamefont {Jelezko}}, \ and\ \bibinfo {author} {\bibfnamefont
  {J.}~\bibnamefont {Wrachtrup}},\ }\href {\doibase 10.1103/PhysRevB.83.081201}
  {\bibfield  {journal} {\bibinfo  {journal} {Phys. Rev. B}\ }\textbf {\bibinfo
  {volume} {83}},\ \bibinfo {pages} {081201} (\bibinfo {year}
  {2011})}\BibitemShut {NoStop}%
\bibitem [{\citenamefont {Siyushev}\ \emph {et~al.}(2010)\citenamefont
  {Siyushev}, \citenamefont {Kaiser}, \citenamefont {Jacques}, \citenamefont
  {Gerhardt}, \citenamefont {Bischof}, \citenamefont {Fedder}, \citenamefont
  {Dodson}, \citenamefont {Markham}, \citenamefont {Twitchen}, \citenamefont
  {Jelezko},\ and\ \citenamefont {Wrachtrup}}]{siyushev_apl_2010}%
  \BibitemOpen
  \bibfield  {author} {\bibinfo {author} {\bibfnamefont {P.}~\bibnamefont
  {Siyushev}}, \bibinfo {author} {\bibfnamefont {F.}~\bibnamefont {Kaiser}},
  \bibinfo {author} {\bibfnamefont {V.}~\bibnamefont {Jacques}}, \bibinfo
  {author} {\bibfnamefont {I.}~\bibnamefont {Gerhardt}}, \bibinfo {author}
  {\bibfnamefont {S.}~\bibnamefont {Bischof}}, \bibinfo {author} {\bibfnamefont
  {H.}~\bibnamefont {Fedder}}, \bibinfo {author} {\bibfnamefont
  {J.}~\bibnamefont {Dodson}}, \bibinfo {author} {\bibfnamefont
  {M.}~\bibnamefont {Markham}}, \bibinfo {author} {\bibfnamefont
  {D.}~\bibnamefont {Twitchen}}, \bibinfo {author} {\bibfnamefont
  {F.}~\bibnamefont {Jelezko}}, \ and\ \bibinfo {author} {\bibfnamefont
  {J.}~\bibnamefont {Wrachtrup}},\ }\href {\doibase 10.1063/1.3519849}
  {\bibfield  {journal} {\bibinfo  {journal} {Applied Physics Letters}\
  }\textbf {\bibinfo {volume} {97}},\ \bibinfo {eid} {241902} (\bibinfo {year}
  {2010})}\BibitemShut {NoStop}%
\bibitem [{\citenamefont {Marseglia}\ \emph {et~al.}(2011)\citenamefont
  {Marseglia}, \citenamefont {Hadden}, \citenamefont {Stanley-Clarke},
  \citenamefont {Harrison}, \citenamefont {Patton}, \citenamefont {Ho},
  \citenamefont {Naydenov}, \citenamefont {Jelezko}, \citenamefont {Meijer},
  \citenamefont {Dolan}, \citenamefont {Smith}, \citenamefont {Rarity},\ and\
  \citenamefont {O'Brien}}]{marseglia_apl_2011}%
  \BibitemOpen
  \bibfield  {author} {\bibinfo {author} {\bibfnamefont {L.}~\bibnamefont
  {Marseglia}}, \bibinfo {author} {\bibfnamefont {J.~P.}\ \bibnamefont
  {Hadden}}, \bibinfo {author} {\bibfnamefont {A.~C.}\ \bibnamefont
  {Stanley-Clarke}}, \bibinfo {author} {\bibfnamefont {J.~P.}\ \bibnamefont
  {Harrison}}, \bibinfo {author} {\bibfnamefont {B.}~\bibnamefont {Patton}},
  \bibinfo {author} {\bibfnamefont {Y.-L.~D.}\ \bibnamefont {Ho}}, \bibinfo
  {author} {\bibfnamefont {B.}~\bibnamefont {Naydenov}}, \bibinfo {author}
  {\bibfnamefont {F.}~\bibnamefont {Jelezko}}, \bibinfo {author} {\bibfnamefont
  {J.}~\bibnamefont {Meijer}}, \bibinfo {author} {\bibfnamefont {P.~R.}\
  \bibnamefont {Dolan}}, \bibinfo {author} {\bibfnamefont {J.~M.}\ \bibnamefont
  {Smith}}, \bibinfo {author} {\bibfnamefont {J.~G.}\ \bibnamefont {Rarity}}, \
  and\ \bibinfo {author} {\bibfnamefont {J.~L.}\ \bibnamefont {O'Brien}},\
  }\href {\doibase 10.1063/1.3573870} {\bibfield  {journal} {\bibinfo
  {journal} {Applied Physics Letters}\ }\textbf {\bibinfo {volume} {98}},\
  \bibinfo {eid} {133107} (\bibinfo {year} {2011})}\BibitemShut {NoStop}%
\bibitem [{\citenamefont {Baba}\ \emph {et~al.}(1999)\citenamefont {Baba},
  \citenamefont {Sasaki}, \citenamefont {Yoshita},\ and\ \citenamefont
  {Akiyama}}]{baba_jap_1999}%
  \BibitemOpen
  \bibfield  {author} {\bibinfo {author} {\bibfnamefont {M.}~\bibnamefont
  {Baba}}, \bibinfo {author} {\bibfnamefont {T.}~\bibnamefont {Sasaki}},
  \bibinfo {author} {\bibfnamefont {M.}~\bibnamefont {Yoshita}}, \ and\
  \bibinfo {author} {\bibfnamefont {H.}~\bibnamefont {Akiyama}},\ }\href
  {http://jap.aip.org/resource/1/japiau/v85/i9/p6923_s1} {\bibfield  {journal}
  {\bibinfo  {journal} {Journal of Applied Physics}\ }\textbf {\bibinfo
  {volume} {85}},\ \bibinfo {pages} {6923} (\bibinfo {year}
  {1999})}\BibitemShut {NoStop}%
\bibitem [{\citenamefont {Bobroff}(1986)}]{Bobroff1986}%
  \BibitemOpen
  \bibfield  {author} {\bibinfo {author} {\bibfnamefont {N.}~\bibnamefont
  {Bobroff}},\ }\href {\doibase http://dx.doi.org/10.1063/1.1138619} {\bibfield
   {journal} {\bibinfo  {journal} {Review of Scientific Instruments}\ }\textbf
  {\bibinfo {volume} {57}},\ \bibinfo {pages} {1152} (\bibinfo {year}
  {1986})}\BibitemShut {NoStop}%
\bibitem [{\citenamefont {Thompson}, \citenamefont {Larson},\ and\
  \citenamefont {Webb}(2002)}]{Thompson2002}%
  \BibitemOpen
  \bibfield  {author} {\bibinfo {author} {\bibfnamefont {R.~E.}\ \bibnamefont
  {Thompson}}, \bibinfo {author} {\bibfnamefont {D.~R.}\ \bibnamefont
  {Larson}}, \ and\ \bibinfo {author} {\bibfnamefont {W.~W.}\ \bibnamefont
  {Webb}},\ }\href {\doibase http://dx.doi.org/10.1016/S0006-3495(02)75618-X}
  {\bibfield  {journal} {\bibinfo  {journal} {Biophysical Journal}\ }\textbf
  {\bibinfo {volume} {82}},\ \bibinfo {pages} {2775 } (\bibinfo {year}
  {2002})}\BibitemShut {NoStop}%
\bibitem [{\citenamefont {Carlsson}(1991)}]{Carlsson1991}%
  \BibitemOpen
  \bibfield  {author} {\bibinfo {author} {\bibfnamefont {K.}~\bibnamefont
  {Carlsson}},\ }\href {\doibase 10.1111/j.1365-2818.1991.tb03169.x} {\bibfield
   {journal} {\bibinfo  {journal} {Journal of Microscopy}\ }\textbf {\bibinfo
  {volume} {163}},\ \bibinfo {pages} {167} (\bibinfo {year}
  {1991})}\BibitemShut {NoStop}%
\bibitem [{\citenamefont {Hell}\ \emph {et~al.}(1993)\citenamefont {Hell},
  \citenamefont {Reiner}, \citenamefont {Cremer},\ and\ \citenamefont
  {Stelzer}}]{Hell1993}%
  \BibitemOpen
  \bibfield  {author} {\bibinfo {author} {\bibfnamefont {S.}~\bibnamefont
  {Hell}}, \bibinfo {author} {\bibfnamefont {G.}~\bibnamefont {Reiner}},
  \bibinfo {author} {\bibfnamefont {C.}~\bibnamefont {Cremer}}, \ and\ \bibinfo
  {author} {\bibfnamefont {E.~H.~K.}\ \bibnamefont {Stelzer}},\ }\href
  {\doibase 10.1111/j.1365-2818.1993.tb03315.x} {\bibfield  {journal} {\bibinfo
   {journal} {Journal of Microscopy}\ }\textbf {\bibinfo {volume} {169}},\
  \bibinfo {pages} {391} (\bibinfo {year} {1993})}\BibitemShut {NoStop}%
\bibitem [{\citenamefont {Visser}, \citenamefont {Oud},\ and\ \citenamefont
  {Brakenhoff}(1992)}]{Visser1992}%
  \BibitemOpen
  \bibfield  {author} {\bibinfo {author} {\bibfnamefont {T.~D.}\ \bibnamefont
  {Visser}}, \bibinfo {author} {\bibfnamefont {J.~L.}\ \bibnamefont {Oud}}, \
  and\ \bibinfo {author} {\bibfnamefont {G.~J.}\ \bibnamefont {Brakenhoff}},\
  }\href@noop {} {\bibfield  {journal} {\bibinfo  {journal} {Optik}\ }\textbf
  {\bibinfo {volume} {90}},\ \bibinfo {pages} {17} (\bibinfo {year}
  {1992})}\BibitemShut {NoStop}%
\bibitem [{\citenamefont {Ohno}\ \emph {et~al.}(2012)\citenamefont {Ohno},
  \citenamefont {{Joseph Heremans}}, \citenamefont {Bassett}, \citenamefont
  {Myers}, \citenamefont {Toyli}, \citenamefont {{Bleszynski Jayich}},
  \citenamefont {Palmstr{\o}m},\ and\ \citenamefont {Awschalom}}]{Ohno2012}%
  \BibitemOpen
  \bibfield  {author} {\bibinfo {author} {\bibfnamefont {K.}~\bibnamefont
  {Ohno}}, \bibinfo {author} {\bibfnamefont {F.}~\bibnamefont {{Joseph
  Heremans}}}, \bibinfo {author} {\bibfnamefont {L.~C.}\ \bibnamefont
  {Bassett}}, \bibinfo {author} {\bibfnamefont {B.~A.}\ \bibnamefont {Myers}},
  \bibinfo {author} {\bibfnamefont {D.~M.}\ \bibnamefont {Toyli}}, \bibinfo
  {author} {\bibfnamefont {A.~C.}\ \bibnamefont {{Bleszynski Jayich}}},
  \bibinfo {author} {\bibfnamefont {C.~J.}\ \bibnamefont {Palmstr{\o}m}}, \
  and\ \bibinfo {author} {\bibfnamefont {D.~D.}\ \bibnamefont {Awschalom}},\
  }\href {\doibase 10.1063/1.4748280} {\bibfield  {journal} {\bibinfo
  {journal} {Applied Physics Letters}\ }\textbf {\bibinfo {volume} {101}},\
  \bibinfo {pages} {082413} (\bibinfo {year} {2012})}\BibitemShut {NoStop}%
\bibitem [{\citenamefont {Widmann}\ \emph {et~al.}(2014)\citenamefont
  {Widmann}, \citenamefont {Lee}, \citenamefont {Rendler}, \citenamefont {Son},
  \citenamefont {Fedder}, \citenamefont {Paik}, \citenamefont {Zhao},
  \citenamefont {Yang}, \citenamefont {Booker}, \citenamefont {Denisenko},
  \citenamefont {Jamali}, \citenamefont {Momenzadeh}, \citenamefont {Ohshima},
  \citenamefont {Gali}, \citenamefont {Janz\'en},\ and\ \citenamefont
  {Wrachtrup}}]{widmann_coherent_2014}%
  \BibitemOpen
  \bibfield  {author} {\bibinfo {author} {\bibfnamefont {M.}~\bibnamefont
  {Widmann}}, \bibinfo {author} {\bibfnamefont {S.-Y.}\ \bibnamefont {Lee}},
  \bibinfo {author} {\bibfnamefont {T.}~\bibnamefont {Rendler}}, \bibinfo
  {author} {\bibfnamefont {N.~T.}\ \bibnamefont {Son}}, \bibinfo {author}
  {\bibfnamefont {H.}~\bibnamefont {Fedder}}, \bibinfo {author} {\bibfnamefont
  {S.}~\bibnamefont {Paik}}, \bibinfo {author} {\bibfnamefont {N.}~\bibnamefont
  {Zhao}}, \bibinfo {author} {\bibfnamefont {S.}~\bibnamefont {Yang}}, \bibinfo
  {author} {\bibfnamefont {I.}~\bibnamefont {Booker}}, \bibinfo {author}
  {\bibfnamefont {A.}~\bibnamefont {Denisenko}}, \bibinfo {author}
  {\bibfnamefont {M.}~\bibnamefont {Jamali}}, \bibinfo {author} {\bibfnamefont
  {S.~A.}\ \bibnamefont {Momenzadeh}}, \bibinfo {author} {\bibfnamefont
  {T.}~\bibnamefont {Ohshima}}, \bibinfo {author} {\bibfnamefont
  {A.}~\bibnamefont {Gali}}, \bibinfo {author} {\bibfnamefont {E.}~\bibnamefont
  {Janz\'en}}, \ and\ \bibinfo {author} {\bibfnamefont {J.}~\bibnamefont
  {Wrachtrup}},\ }\href {http://arxiv.org/abs/1407.0180} {\bibfield  {journal}
  {\bibinfo  {journal} {{arXiv}:1407.0180 [cond-mat]}\ } (\bibinfo {year}
  {2014})},\ \bibinfo {note} {{arXiv}: 1407.0180}\BibitemShut {NoStop}%
\bibitem [{\citenamefont {Kitson}\ \emph {et~al.}(1998)\citenamefont {Kitson},
  \citenamefont {Jonsson}, \citenamefont {Rarity},\ and\ \citenamefont
  {Tapster}}]{kitson_pra_1998}%
  \BibitemOpen
  \bibfield  {author} {\bibinfo {author} {\bibfnamefont {S.~C.}\ \bibnamefont
  {Kitson}}, \bibinfo {author} {\bibfnamefont {P.}~\bibnamefont {Jonsson}},
  \bibinfo {author} {\bibfnamefont {J.~G.}\ \bibnamefont {Rarity}}, \ and\
  \bibinfo {author} {\bibfnamefont {P.~R.}\ \bibnamefont {Tapster}},\ }\href
  {\doibase 10.1103/PhysRevA.58.620} {\bibfield  {journal} {\bibinfo  {journal}
  {Phys. Rev. A}\ }\textbf {\bibinfo {volume} {58}},\ \bibinfo {pages} {620}
  (\bibinfo {year} {1998})}\BibitemShut {NoStop}%
\bibitem [{\citenamefont {Beha}\ \emph {et~al.}(2012)\citenamefont {Beha},
  \citenamefont {Batalov}, \citenamefont {Manson}, \citenamefont
  {Bratschitsch},\ and\ \citenamefont {Leitenstorfer}}]{Beha2012}%
  \BibitemOpen
  \bibfield  {author} {\bibinfo {author} {\bibfnamefont {K.}~\bibnamefont
  {Beha}}, \bibinfo {author} {\bibfnamefont {a.}~\bibnamefont {Batalov}},
  \bibinfo {author} {\bibfnamefont {N.~B.}\ \bibnamefont {Manson}}, \bibinfo
  {author} {\bibfnamefont {R.}~\bibnamefont {Bratschitsch}}, \ and\ \bibinfo
  {author} {\bibfnamefont {a.}~\bibnamefont {Leitenstorfer}},\ }\href {\doibase
  10.1103/PhysRevLett.109.097404} {\bibfield  {journal} {\bibinfo  {journal}
  {Physical Review Letters}\ }\textbf {\bibinfo {volume} {109}},\ \bibinfo
  {pages} {097404} (\bibinfo {year} {2012})}\BibitemShut {NoStop}%
\bibitem [{\citenamefont {Han}\ \emph {et~al.}(2012)\citenamefont {Han},
  \citenamefont {Wildanger}, \citenamefont {Rittweger}, \citenamefont {Meijer},
  \citenamefont {Pezzagna}, \citenamefont {Hell},\ and\ \citenamefont
  {Eggeling}}]{Han2012}%
  \BibitemOpen
  \bibfield  {author} {\bibinfo {author} {\bibfnamefont {K.~Y.}\ \bibnamefont
  {Han}}, \bibinfo {author} {\bibfnamefont {D.}~\bibnamefont {Wildanger}},
  \bibinfo {author} {\bibfnamefont {E.}~\bibnamefont {Rittweger}}, \bibinfo
  {author} {\bibfnamefont {J.}~\bibnamefont {Meijer}}, \bibinfo {author}
  {\bibfnamefont {S.}~\bibnamefont {Pezzagna}}, \bibinfo {author}
  {\bibfnamefont {S.~W.}\ \bibnamefont {Hell}}, \ and\ \bibinfo {author}
  {\bibfnamefont {C.}~\bibnamefont {Eggeling}},\ }\href {\doibase
  10.1088/1367-2630/14/12/123002} {\bibfield  {journal} {\bibinfo  {journal}
  {New Journal of Physics}\ }\textbf {\bibinfo {volume} {14}},\ \bibinfo
  {pages} {123002} (\bibinfo {year} {2012})}\BibitemShut {NoStop}%
\bibitem [{\citenamefont {Aharonovich}\ \emph {et~al.}(2010)\citenamefont
  {Aharonovich}, \citenamefont {Castelletto}, \citenamefont {Simpson},
  \citenamefont {Greentree},\ and\ \citenamefont {Prawer}}]{Aharonovich2010}%
  \BibitemOpen
  \bibfield  {author} {\bibinfo {author} {\bibfnamefont {I.}~\bibnamefont
  {Aharonovich}}, \bibinfo {author} {\bibfnamefont {S.}~\bibnamefont
  {Castelletto}}, \bibinfo {author} {\bibfnamefont {D.~a.}\ \bibnamefont
  {Simpson}}, \bibinfo {author} {\bibfnamefont {a.~D.}\ \bibnamefont
  {Greentree}}, \ and\ \bibinfo {author} {\bibfnamefont {S.}~\bibnamefont
  {Prawer}},\ }\href {\doibase 10.1103/PhysRevA.81.043813} {\bibfield
  {journal} {\bibinfo  {journal} {Physical Review A}\ }\textbf {\bibinfo
  {volume} {81}},\ \bibinfo {pages} {1} (\bibinfo {year} {2010})}\BibitemShut
  {NoStop}%
\bibitem [{\citenamefont {Neu}, \citenamefont {Agio},\ and\ \citenamefont
  {Becher}(2012)}]{Neu2012}%
  \BibitemOpen
  \bibfield  {author} {\bibinfo {author} {\bibfnamefont {E.}~\bibnamefont
  {Neu}}, \bibinfo {author} {\bibfnamefont {M.}~\bibnamefont {Agio}}, \ and\
  \bibinfo {author} {\bibfnamefont {C.}~\bibnamefont {Becher}},\ }\href
  {http://arxiv.org/abs/1107.0502 http://www.ncbi.nlm.nih.gov/pubmed/23037048}
  {\bibfield  {journal} {\bibinfo  {journal} {Optics express}\ }\textbf
  {\bibinfo {volume} {20}},\ \bibinfo {pages} {19956} (\bibinfo {year}
  {2012})},\ \Eprint {http://arxiv.org/abs/1107.0502} {arXiv:1107.0502}
  \BibitemShut {NoStop}%
\end{thebibliography}%
\end{document}